\documentclass[journal,a4paper]{IEEEtran} 
\usepackage{graphicx}

\newcommand{\onetom}{1,\cdots,m}

\begin{document}
%
\title{ Pinning complex Networks by A Single
Controller}

\author{Tianping Chen, Xiwei Liu and Wenlian Lu
\thanks{It is supported
  by National Science Foundation of China 69982003 and 60074005, and also
  supported by Graduate
  Innovation Foundation of Fudan University.}
\thanks{These authors are with Lab. of
Nonlinear Mathematics Science, Institute of Mathematics, Fudan
University, Shanghai, 200433, P.R.China. The third author is also
with Max Planck Institute for Mathematics in the Sciences,
Leipzig, Germany. Corresponding author: Tianping Chen.
Email:tchen@fudan.edu.cn}}

\maketitle

{\bf Abstract} In this short paper, we point out that a single
local stability controller can pin a linear or nonlinear coupled
complex network to a specified solution (or an equilibrium) of the
coupled complex network. A rigorous mathematical proof is given,
too.
\\


\IEEEpeerreviewmaketitle
\section{Introduction}
\quad Many natural and man-made systems, such as neural systems,
social systems, WWW, food webs, electrical power grids, etc., can
all be described by graphs. In such a graph, every node represents
an individual element of the system, while edges represent
relations between nodes. These graphs are called complex networks.
For decades, complex networks have been focused on by scientists
from various fields, for instance, sociology, biology, mathematics
and physics,etc.

Linearly Coupled Ordinary Differential Equations (LCODEs) are a
large class of dynamical systems with continuous time and state,
as well as discrete space to describe coupled oscillators.

In general, the LCODEs can be described as follows:
\begin{eqnarray*}
\frac{d x_{i}(t)}{dt}=f(x_{i}(t),t)+\sum\limits_{j=1,j\ne
i}^{m}a_{ij}[x_{j}(t)-x_{i}(t)],\\\quad i=\onetom
\end{eqnarray*}
where $x_{i}(t)=[x_{i}^{1}(t),\cdots,x_{i}^{n}(t)]^{\top}\in
R^{n}$ is the state variable of the $i-th$ node, $t\in
[0,+\infty)$ is a continuous time,
$f:R\times[0,+\infty)\rightarrow R^{n}$ is continuous map,
$a_{ij}\geq 0$ for $i,j=1,\cdots,m, $ $i\ne j$.

Letting $a_{ii}=-\sum\limits_{j=1,j\ne i}^{m}a_{ij}$, the equation
above can be rewritten as follows:
\begin{eqnarray}
\frac{d x_{i}(t)}{dt}=f(x_{i}(t),t)+\sum\limits_{j=1}^{m}a_{ij}
x_{j}(t)\quad i=\onetom\label{synn}
\end{eqnarray}
where $A=(a_{ij})\in R^{m\times m}$ is the coupling matrix with
zero-sum rows and $a_{ij}\ge 0$, for  $i\ne j$,  which is
determined by the topological structure of the LCODEs.

Today, in science and technology , ``synchronization''  appears in
a wide range of real systems.  In theoretical fields, there are
various kinds of concepts of synchronization. For example, phase
synchronization, imperfect synchronization, lag synchronization,
and almost synchronization, etc. In this paper, the concept of
complete synchronization is considered, which can be
mathematically defined as follows: If a coupled system composed of
$m$ sub-systems
\begin{eqnarray*}
\dot{x}_{i}=f_{i}(x_{1},\cdots,x_{m},t),\quad for~i=\onetom
\end{eqnarray*}
satisfies
$\lim\limits_{t\rightarrow\infty}\|x_{j}(t)-x_{i}(t)\|=0$, for all
$i,j=\onetom$, then the coupled system is said to be completely
synchronized, for simplicity, synchronized.

The synchronization of the LCODEs has attracted much attention of
researchers in different fields. In \cite{Pec1,Wang1}, the local
stability of the synchronization manifold was studied via
linearization method named ``master stability function''. In
\cite{Wu}, a distance was defined between the state of the coupled
system and synchronization manifold. Using this distance, a
methodology was proposed to discuss global convergence for
complete regular coupling configuration (also see \cite{Chen1}).

In the paper \cite{Chen2}, based on the geometric structure of the
synchronization manifold, a thorough theoretical analysis  of the
synchronization of LCODEs is given. Moreover, it was pointed out
that the synchronization of the coupled system only means that
$\lim\limits_{t\rightarrow\infty}\|x_{j}(t)-x_{i}(t)\|=0$, for all
$i,j=\onetom$. It does not mean that
$\lim\limits_{t\rightarrow\infty}x_{j}(t)=s(t)$ for some $s(t)$
satisfying $\dot{s}(t)=f(s(t))$. Even if all the differences
$\|x_{j}(0)-s(0)\|$ are very small, it still hold that
$\lim\limits_{t\rightarrow\infty}\|x_{j}(t)-s(t)\|\ne 0$.

However, in some cases, it is needed that all the states of the
the nodes synchronize to some special solution $s(t)$ of the
system $\dot{s}(t)=f(s(t))$ or some equilibrium point $s^{*}$
satisfying $f(s^{*})=0$. Moreover, it often happens that due to
some reasons, some nodes can not receive controller. thus, it is
natural to raise the question: can we realize synchronize all
states $x_{i}(t)$ to $s(t)$ only add control to some nodes.

In the paper \cite{Li}, the authors discussed how to pin a complex
dynamical networks, where the coupling matrix is symmetric, to its
equilibrium by adding some controllers. They pointed out that
significantly less (but do not specify how many) local stable
controllers can pin a complex dynamical networks to its
equilibrium. Here, we rigorously  prove that only a single
controller can achieve this goal.

\section{Pin a complex network with linear coupling}\quad
Suppose that the complex network is
\begin{equation}
\frac{dx_i(t)}{dt}=f(x_i(t),t)+c\sum\limits_{j=1}^ma_{ij}x_j(t)\quad
i=1,\cdots,m
\end{equation}
where $x_i\in R^n$, $a_{ij}\geq0, i\not=j$ and
$\sum\limits_{j=1}^{m}a_{ij}=0$, for $i=1,2,\cdots,m$. $s(t)$ is a
solution of the uncoupled system
\begin{equation}
\dot{s}(t)=f(s(t))
\end{equation}

We prove that if $\varepsilon>0$ and $c$ is large enough. The
following coupled network with a single controller
\begin{eqnarray}
\frac{dx_1(t)}{dt}&=&f(x_1(t),t)+c\sum\limits_{j=1}^ma_{1j}x_j(t)-c\varepsilon(x_{1}(t)-s(t)),\nonumber\\
\frac{dx_i(t)}{dt}&=&f(x_i(t),t)+c\sum\limits_{j=1}^ma_{ij}x_j(t),~
i=2,\cdots,m \label{pin}
\end{eqnarray}
can pin the complex dynamical network $(1)$ to $s(t)$.

Denote $\delta x_i(t)=x_i(t)-s(t)$, then the system $(1)$ can be
rewritten as:
\begin{eqnarray}
\frac{d\delta
x_i(t)}{dt}=f(x_i(t),t)-f(s(t))+c\sum\limits_{j=1}^m{a}_{ij}\delta
x_j(t),\nonumber\\\quad i=1,\cdots,m
\end{eqnarray}
and the network with a single controller $(\ref{pin})$ is written
as
\begin{eqnarray}
\frac{d\delta
x_i(t)}{dt}=f(x_i(t),t)-f(s(t))+c\sum\limits_{j=1}^m\tilde{a}_{ij}\delta
x_j(t),\nonumber\\~ i=1,\cdots,m
\end{eqnarray}
where $\tilde{a}_{11}=a_{11}-\varepsilon$, $\varepsilon>0$ and
$\tilde{a}_{ij}=a_{ij}$ otherwise.

At first, we prove the following simple proposition.

{\bf Proposition 1}\quad If  $A=(a_{ij})_{i,j=1}^{m}$ is a matrix
satisfying $a_{ij}=a_{ji}\geq0,$  if $ i\not=j$, and
$\sum\limits_{j=1}^{m}a_{ij}=0$, for $i=1,2,\cdots,m$. Then,  all
eigenvalues of the matrix
\[\tilde{A}=\left(\begin{array}{cccc}a_{11}-\varepsilon&a_{12}&\cdots&a_{1m}\\
a_{21}&a_{22}&\cdots&a_{2m}\\\vdots&\vdots&\ddots&\vdots\\
a_{m1}&a_{m2}&\cdots&a_{mm}\end{array}\right)\] are negative.

{\bf Proof}\quad Suppose that $\lambda$ is an eigenvalue of
$\tilde{A}$, $v=[v_{1},\cdots,v_{m}]^{T}$ is the corresponding
eigenvector, and $|v_{k}|=max_{j=1,\cdots,m}|v_{j}|$. It is clear
that if $v$ is an eigenvector, then $-v$ is also an eigenvector.
thus, without loss of generality, we can assume that $v_{k}>0$ and
$v_{k}=max_{j=1,\cdots,m}|v_{j}|$.

If $k=1$. Then
\begin{eqnarray*}
\sum_{j=1}^{m}\tilde{a}_{1j}v_{j}&=&-\varepsilon
v_{1}+\sum_{j=1}^{m}a_{1j}v_{j}\\&\le &-\varepsilon
v_{1}+\sum_{j=1}^{m}a_{1j}|v_{j}|< -\varepsilon v_{1}<0
\end{eqnarray*}
which means $\lambda<0$.

Instead, if $k>1$. Then,
\begin{eqnarray*}
\lambda v_{k}=\sum_{j=1}^{m}\tilde{a}_{kj}v_{j}\le
\tilde{a}_{kk}v_{k}+\sum_{j\ne k}^{m}\tilde{a}_{kj}|v_{j}|\le 0
\end{eqnarray*}
which means $\lambda\le 0$. If $\lambda=0$, then,
$v=[v_{k}\cdots,v_{k}]^{T}$. However, this is impossible. For
\begin{eqnarray*}
\sum_{j=1}^{m}\tilde{a}_{1j}v_{j}=-\epsilon v_{k}< 0
\end{eqnarray*}
Therefore, $\lambda<0$. The proposition is proved.

\subsection{Pin a complex network with linear coupling and symmetric coupling matrix}
In this section, we investigate the complex networks, where the
coupling is linear and the coupling matrix is symmetric.

With the simple proposition given above, we prove two theorems.
Theorem 1 addresses local synchronization. Theorem 2 addresses
global synchronization.

\qquad Let $\delta x(t)=[\delta x_1(t),\cdots,\delta x_m(t)]$.
Differentiating  along $s(t)$ gives
\begin{equation}
\frac{d\delta x(t)}{dt}=D(f(s(t)))\delta x(t)+c\delta
x(t)\tilde{A}^{\top}
\end{equation}

Let $\tilde{A}^T=WJW^{\top}$ be the eigenvalue decomposition of
$\tilde{A}$, where $J=diag\{\lambda_1,\cdots,\lambda_m\}$,
$0>\lambda_1>\cdots>\lambda_m$, and $\delta y(t)=\delta x(t)W$.
Then we have
\begin{eqnarray}
\frac{d\delta y_{k}(t)}{dt}&=&[Df(s(t))+c\lambda_kI]\delta
y_{k}(t)\end{eqnarray}

{\bf Theorem 1.}\quad Let $\mu_{i}(t)$, $i=1,\cdots,m$, are the
eigenvalues of the matrix $\frac{1}{2}(Df(s(t))+Df^{T}(s(t)))$,
$\mu(t)=\max_{i=1,\cdots,m}\mu_{i}(t)$. If $\mu(t)<-c\lambda_{1}$
for all $t>0$. Then, the coupled system with a controller (4) can
be locally exponentially synchronized to $s(t)$.

{\bf Proof:}\quad It is easy to see that
\begin{eqnarray}
&&\frac{1}{2}\frac{d\{\delta y_{k}^{\top}(t)\delta
y_{k}(t)\}}{dt}=\delta y_{k}^{\top}(t)[Df(s(t))+c\lambda_kI]\delta
y_{k}(t)\nonumber\\&=& \delta
y_{k}^{\top}(t)[\frac{1}{2}(Df(s(t))+Df^{T}(s(t)))+c\lambda_kI]\delta
y_{k}(t)
\end{eqnarray}
Under $\mu(t)<-c\lambda_{1}$ for all $t>0$, we have
\begin{eqnarray}
\frac{1}{2}\frac{d\{\delta y_{k}^{\top}(t)\delta
y_{k}(t)\}}{dt}\le \delta
y_{k}^{\top}(t)[\mu(t)-c\lambda_{1}]\delta y_{k}(t)
\end{eqnarray}
which means that $\delta y_{k}^{\top}(t)\delta
y_{k}(t)=O(e^{(\mu(t)-c\lambda_{1})t})$. Theorem 1 is proved.

{\bf Theorem 2.}\quad Suppose $0>\lambda_1>\cdots>\lambda_m$ are
the eigenvalues of $\tilde{A}$. If there are  positive diagonal
matrices $P=diag\{p_1,\cdots,p_{n}\}$,
$\Delta=diag\{\Delta_1,\cdots,\Delta_n\}$ and a constant $\eta>0$,
such that
\begin{equation}(x-y)^{\top}P(f(x,t)-\Delta x-f(y,t)+\Delta y)
\leq -\eta (x-y)^{\top}(x-y)
\end{equation}
and $\Delta_k+c\lambda_1<0$ for $k=1,\cdots,n$. Then, the
controlled system (4) is globally exponentially synchronized to
$s(t)$.

{\bf Proof:}\quad Define a Lyapunov function as
\[V(t)=\frac{1}{2}\sum\limits_{i=1}^m\delta x_i(t)^{\top}P\delta
x_i(t)\] Denote
$\delta\tilde{x}^k(t)=[\delta{x}^k_1(t),\cdots,\delta{x}^k_m(t)]^{\top}$.
Then, we have
\begin{eqnarray*}
&&\frac{dV(t)}{dt}=\sum\limits_{i=1}^m\delta
x_i(t)^{\top}Pf(x_i(t),t)\delta
x_j(t)\\&-&\sum\limits_{i=1}^m\delta x_i(t)^{\top}Pf(s(t))\delta
x_j(t)\\&+&\sum\limits_{i=1}^m\delta x_i(t)^{\top}
Pc\sum\limits_{j=1}^m\tilde{a}_{ij}\delta
x_j(t)\\&=&\sum\limits_{i=1}^m\delta
x_i(t)^{\top}\bigg[P(f(x_i(t),t)-f(s(t)))-\Delta \delta
x_i(t)\bigg]\\&+&\sum\limits_{i=1}^m\delta
x_i(t)^TP\bigg[c\sum\limits_{j=1}^m\tilde{a}_{ij}\delta
x_j(t)+\Delta \delta
x_i(t)\bigg]\\
&\leq&-\eta\sum\limits_{i=1}^m\delta x_i(t)^{\top}\delta
x_i(t)\\&+&\sum\limits_{i=1}^m\delta
x_i(t)^{\top}P\bigg[c\sum\limits_{j=1}^m\tilde{a}_{ij}\delta
x_j(t)+\Delta \delta x_i(t)\bigg]
\\&=&-\eta\sum\limits_{i=1}^m\delta x_i(t)^{\top}\delta
x_i(t)\\
&+&\sum\limits_{k=1}^n p_k\delta\tilde{x}^{k}(t)^{\top}(c
\tilde{A}+\Delta_kI)\delta \tilde{x}^{k}(t)
\end{eqnarray*}
Because $c \tilde{A}+\Delta_kI<0$, we have
\begin{eqnarray*}
&&\frac{dV(t)}{dt}\le -\eta\sum\limits_{i=1}^m\delta
x_i(t)^{\top}\delta x_i(t)\leq -\eta\frac{V(t)}{\min_{i}p_i}
\end{eqnarray*}
 Therefore
\begin{eqnarray*}
V(t)=O(e^{\frac{-\eta t}{\min_{i}p_i}})
\end{eqnarray*}
Theorem 2 is proved completely.

{\bf Remark 1}\quad It is clear that if $c$ is large enough, then
the coupled network with a single controller can pin the complex
network to a solution $s(t)$ of the uncoupled system.

\subsection{Pin a complex network with nonlinear coupling}\quad
In this section, we discuss how to pin a complex network with
nonlinear coupling. In this case, the coupled system (\ref{pin})
takes the following form
\begin{eqnarray}
\frac{dx_1(t)}{dt}=f(x_1(t),t)+c\sum\limits_{j=1}^ma_{1j}g(x_j(t))
\nonumber\\-c\varepsilon(g(x_{1}(t))-g(s(t))),\nonumber\\
\frac{dx_i(t)}{dt}=f(x_i(t),t)+c\sum\limits_{j=1}^ma_{ij}g(x_j(t))
\quad i=2,\cdots,m \label{pin1}
\end{eqnarray}
where
$g(x_{i}(t))=[g(x_{i}^{1}(t)),\cdots,g(x_{i}^{n}(t))]^{\top}$ and
$g(x)$ is a monotone increasing function.

In the following, we will prove that the coupled complex network
with a single controller (\ref{pin1}) can pin it to a specified
solution $\dot{s}(t)=f(s(t),t)$, too. In particular, we prove

{\bf Theorem 3.}\quad Suppose $0>\lambda_1>\cdots>\lambda_m$ are
the eigenvalues of $\tilde{A}$,
$\frac{g(u)-g(v)}{u-v}\geq\underline{\alpha}> 0$. If there are
positive diagonal matrices $P=diag\{p_1,\cdots,p_{n}\}$,
$\Delta=diag\{\Delta_1,\cdots,\Delta_n\}$ and a constant $\eta>0$,
such that
\begin{equation}(x-y)^{\top}P(f(x,t)-\Delta x-f(y,t)+\Delta y)
\leq -\eta (x-y)^{\top}(x-y)
\end{equation}
and $\Delta_k+\underline{\alpha}c\lambda_1<0$ for $k=1,\cdots,n$.
Then, the controlled system (\ref{pin1}) is globally exponentially
synchronized to $s(t)$.

{\bf Proof}\quad Along with
$g(x_{i})=[g(x_{i}^{1}),\cdots,g(x_{i}^{n})]^{\top}$,
$i=1,\cdots,m$, we denote
$\tilde{g}(x^{k})=[g(x^{k}_{1}),\cdots,g(x^{k}_{m})]^{\top}$,
$k=1,\cdots,n$.

We use the same Lyapunov function
\[V(t)=\frac{1}{2}\sum\limits_{i=1}^m\delta x_i(t)^{\top}P\delta x_i(t)\]
In this case, we have
\begin{eqnarray*}
&&\frac{d V(t)}{dt}=\sum\limits_{i=1}^{m}\delta{x}_{i}^{\top}(t)
P\frac{d\delta{x}_{i}(t)}{dt}\\
&=&\sum\limits_{i=1}^{m}\delta{x}_{i}^{\top}(t)P\bigg[f(x^{i}(t),t)-f(
s(t),t)\\& +&c\sum\limits_{j=1}^{m}a_{ij}
g(x_{j}(t))\bigg]\\
&&-c\varepsilon[x_{1}(t)-s(t)]^{\top}P[g({x}_{1}(t))-g(s(t))]\\
&\le& -\eta \sum\limits_{i=1}^{m}\delta{x}_{i}^{\top}(t)\delta
x^{i}(t)+\sum\limits_{i=1}^{m}\delta{x}_{i}^{\top}(t)P\bigg[
\Delta \delta{x}_{i}(t)\\
&&+c\sum\limits_{j=1}^{m}a_{ij}
g(x_{j}(t))\bigg]\\
&&-c\varepsilon[x_{1}(t)-s(t)]^{\top}P\underline{\alpha}[x_{1}(t)-s(t)]
\end{eqnarray*}
By the property of the matrix $A$, it is easy to verify that for
$u=[u_{1},\cdots,u_{m}]^{\top}$, $v=[v_{1},\cdots,v_{m}]^{\top}$,
\begin{eqnarray}
u^{\top}Av=\sum_{i,j=1}^{m}u_{i}a_{ij}v_{j}=\sum_{j>i}a_{ij}(u_{i}-u_{j})(v_{i}-v_{j})
\end{eqnarray}
combining with $\frac{g(u)-g(v)}{u-v}\geq\underline{\alpha}> 0$,
we have
\begin{eqnarray*}
&&\sum\limits_{i=1}^{m}\delta{x}_{i}^{\top}(t)P\sum\limits_{j=1}^{m}a_{ij}
g(x_{j}(t))\\&=& \sum\limits_{k=1}^n
p_k\delta\tilde{x}^{k}(t)^{\top}
A\delta \tilde{g}(x^{k}(t))\\
&=& \sum\limits_{k=1}^n
p_k\sum_{j>i}a_{ij}(x_{i}^{k}(t)-x_{j}^{k}(t))(g(x_{i}^{k}(t))-g(x_{j}^{k}(t)))\\
&\le& -\underline{\alpha}\sum\limits_{k=1}^n
p_k\sum_{j>i}a_{ij}(x_{i}^{k}(t)-x_{j}^{k}(t))(x_{i}^{k}(t)-x_{j}^{k}(t))\\
&=& -\underline{\alpha}\sum_{k=1}^n
p_k\delta\tilde{x}^{k}(t)^{\top} A\delta\tilde{x}^{k}(t)
\end{eqnarray*}
Therefore,
\begin{eqnarray*}
&&\frac{d V(t)}{dt} \le -\eta
\sum\limits_{i=1}^{m}\delta{x}_{i}^{\top}(t)\delta
x^{i}(t)\\&-&\sum_{k=1}^n p_k\delta\tilde{x}^{k}(t)^{\top}
[\Delta_{k}I_{m}+c\underline{\alpha}\tilde{A}]\delta\tilde{x}^{k}(t)\end{eqnarray*}

Because $\Delta_k+\underline{\alpha}c\lambda_1<0$, we have
\begin{eqnarray*}
&&\frac{d V(t)}{dt}\le -\eta\frac{V(t)}{\min_{i}p_i}
\end{eqnarray*}
and
\begin{eqnarray*}
V(t)=O(e^{\frac{-\eta t}{\min_{i}p_i}})
\end{eqnarray*}
Theorem 3 is proved.

\subsection{Pin a complex network with asymmetric coupling
matrix}\quad In practice, indirect graphs are small part. Most of
the graphs are direct graphs. It means the coupling matrix is
asymmetric. Therefore, we must investigate pining the complex
networks, in which the coupling matrix is asymmetric. This is the
issue we investigate in this section.

Consider the system
\begin{eqnarray*}
\frac{dx_1(t)}{dt}=f(x_1(t),t)+c\sum\limits_{j=1}^ma_{1j}x_j(t)-c\varepsilon(x_{1}(t)-s(t)),
\end{eqnarray*}
\begin{eqnarray}
\frac{dx_i(t)}{dt}=f(x_i(t),t)+c\sum\limits_{j=1}^ma_{ij}x_j(t),~
i=2,\cdots,m
\end{eqnarray}
where the coupling matrix $A$ is asymmetric but satisfies zero row
sum $\sum_{j=1}^{m}a_{ij}=0$.

Let $[\xi_{1},\cdots,\xi_{m}]^{T}$ be the left eigenvalue of the
matrix $A$. It is well know that if $A$ is irreducible, all
$\xi_{i}>0$, $i=1,\cdots,m$.

Define $\Xi=diag[\xi_{1},\cdots,\xi_{m}]$. It is easy to verify
that $\Xi A+A^{T}\Xi$ is a symmetric matrix and zero row sum.
Therefore, by the Proposition 1,
$\Xi\tilde{A}+\tilde{A}^{T}\Xi<0$.

{\bf Theorem 4.}\quad Suppose that $A$ is irreducible,
$0>\mu_1>\cdots>\mu_m$ are the eigenvalues of
$(\Xi\tilde{A}+\tilde{A}^{T}\Xi)/2$. If there are positive
diagonal matrices $P=diag\{p_1,\cdots,p_{n}\}$,
$\Delta=diag\{\Delta_1,\cdots,\Delta_n\}$ and a constant $\eta>0$,
such that
\begin{equation}(x-y)^{\top}P(f(x,t)-\Delta x-f(y,t)+\Delta y)
\leq -\eta (x-y)^{\top}(x-y)
\end{equation}
and $\Delta_k\max\limits_{i=1,\cdots,m}\xi_{i}+c\mu_1<0$ for
$k=1,\cdots,n$. Then, the controlled system (4) is globally
exponentially synchronized to $s(t)$.

{\bf Proof:}\quad In this case, define a new Lyapunov function as
\begin{eqnarray*}
V(\delta{x})=\frac{1}{2}\sum_{i=1}^{m}\xi_{i}\delta{x}_{i}^{T}P\delta{x}_{i}
\end{eqnarray*}

Differentiating $V(\delta{x})$ $f(x,t)\in QUAD(\Delta,P)$, we have
\begin{eqnarray*}
&&\frac{d
V(\delta{x})}{dt}=\sum\limits_{i=1}^{m}\xi_{i}\delta{x}_{i}^{T}(t)P\frac{d
\delta x_{i}(t) }{dt}\nonumber\\
&=&\sum\limits_{i=1}^{m}\xi_{i}\delta{x}_{i}^{T}(t)P\bigg[f(x_{i}(t),t)-f(s(t),t)\bigg]\\
&+&c\sum\limits_{i=1}^{m}\xi_{i}\delta{x}_{i}^{T}(t)P\sum\limits_{j=1}^{m}\tilde{a}_{ij}
\delta x_{j}(t)\nonumber\\
&=&\sum\limits_{i=1}^{m}\xi_{i}\delta{x}_{i}^{T}(t)P\bigg[(f(x_{i}(t),t)-\Delta{x}_{i}(t)
)-(f(s(t),t)\\&-&\Delta
s(t))+c\sum\limits_{j=1}^{m}\tilde{a}_{ij}\delta x_{j}(t)+\Delta
\delta x_{i}(t) \bigg]\nonumber\\
&\le&-\eta \sum\limits_{i=1}^{m}\xi_{i}\delta{x}_{i}^{T}(t)\delta
x^{i}(t)\\
& +&\sum\limits_{i=1}^{m}\xi_{i}\delta{x}_{i}^{T}(t)P
\bigg[c\sum\limits_{j=1}^{m}\tilde{a}_{ij}\delta x_{j}(t)+\Delta
\delta x_{i}(t) \bigg]\nonumber\\
&\le&-\eta
V(\delta{x})+\sum\limits_{j=1}^{n}p_{j}\delta\tilde{x}^{j{\top}}(t)\Xi(c\tilde{A}+\Delta_{k}E_{m})
\delta\tilde{x}^{j}(t)\nonumber\\
&=&-\eta\frac{V(t)}{\min_{i}p_i}
+\frac{1}{2}\sum\limits_{j=1}^{n}p_{j}\delta\tilde{x}^{j{\top}}(t)
\bigg[\Xi(c\tilde{A}+\Delta_{k}E_{m})\bigg]^{s}\\
&&\times\delta\tilde{x}^{j}(t)\le-\eta\frac{V(t)}{\min_{i}p_i}
\end{eqnarray*}
For under the assumption $\Delta_j+c\mu_1<0$, we have \newline
$\bigg[\Xi(c\tilde{A}+\Delta_{k}E_{m})\bigg]^{s}$ is negative
definite for $k=1,\cdots,n$.

Therefore
\begin{eqnarray*}
V(t)=O(e^{-\frac{\eta t}{\min_{i}p_i}})
\end{eqnarray*}
Theorem 4 is proved completely.

In the following, we remove assumption that $A$ is irreducible. In
this case, we assume
\begin{eqnarray}
A=\left[\begin{array}{cccc}A_{11}&0&\cdots&0\\
A_{21}&A_{22}&\cdots&0\\
\vdots&\ddots&\vdots&\vdots\\
A_{p1}&A_{p2}&\cdots&A_{pp}\end{array}\right] \label{redu}
\end{eqnarray}
where   $A_{qq}\in R^{m_{q},m_{q}}$, $q=1,\cdots,p$,  are
irreducible. And, for each $q$, there exists $q>k$ such that
$A_{qk}\ne 0$.

It is easy to see that if we add a single controller
$-\epsilon(x_{1}(t)-s(t)$ to the node $x_{1}(t)$. Then, by
previous arguments, we conclude that the subsystem
\begin{eqnarray*}
\frac{dx_1(t)}{dt}=f(x_1(t),t)+c\sum\limits_{j=1}^{m_{1}}a_{1j}x_j(t)
-c\varepsilon(x_{1}(t)-s(t)),
\end{eqnarray*}
\begin{eqnarray}
\frac{dx_i(t)}{dt}=f(x_i(t),t)+c\sum\limits_{j=1}^{m_{1}}a_{ij}x_j(t),~
i=2,\cdots,m
\end{eqnarray}
pins $x_{1}(t),\cdots,x_{m_{1}}$ to $s(t)$.

\begin{eqnarray*}
\frac{dx_i(t)}{dt}=f(x_i(t),t)+c\sum\limits_{j=1}^{m_{2}}a_{ij}x_j(t),~
i=m_{1}+1,\cdots,m_{2}
\end{eqnarray*}
can be written as
\begin{eqnarray*}
\frac{d\delta
x_i(t)}{dt}=f(x_i(t),t)-f(s(t),t)+c\sum\limits_{j=m_{1}+1}^{m_{2}}a_{ij}\delta
x_j(t)+O(e^{-\eta t})
\end{eqnarray*}

Because $A_{21}\ne 0$. Then, in $A_{22}$, there exists at least
one row $i_{2}$, such that
\begin{eqnarray}
a_{i_{2}i_{2}}>\sum_{j=m_{1}+1}^{m_{2}}a_{i_{2}j}x_j(t)
\end{eqnarray}
Therefore, similar to the proposition, we conclude that all
eigenvalues of $A_{22}$ are negative. By the similar arguments in
the proof of theorem 4, we can pin $x_{i}(t)$,
$i=m_{1}+1,\cdots,m_{2}$, to $s(t)$.

By induction, we prove that if we add a controller to the master
subsystem corresponding to the sub-matrix $A_{11}$, then we can
ping the complex network to $s(t)$ even if the coupling matrix in
reducible.

\section{Simulation}\quad
In this section, we give some numerical examples to verify the
theorem given in previous section.

We consider the Chua's circuit
\begin{eqnarray} \left \{
\begin{array}{l}
\frac{dx_{1}}{dt}=k[x_{2}-h(x_{1})]\\
\frac{dx_{2}}{dt}=x_{1}-x_{2}+x_{3}\\
\frac{dx_{3}}{dt}=-lx_{2}
\end{array}
\right. \label{Chua}
\end{eqnarray}
where $h(x)=\frac{2}{7}x-\frac{3}{14}[|x+1|-|x-1|]$, $k=9$ and
$l=14\frac{2}{7}$. With these parameters, the system has a
double-scroll chaotic attractor, as shown in Figure 1.

\begin{figure}
\centering
\includegraphics[width=2.5in]{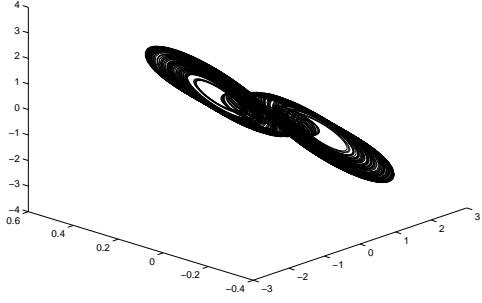}

\caption{Chaotic Behavior of the Chua's circuit.}
\begin{center} \label{fig_sim}
\end{center}
\end{figure}

As the coupled system, we consider three globally coupled Chua's
circuits
\begin{eqnarray}
\frac{dx^i(t)}{dt}=f(x_i(t),t)+c\sum\limits_{j=1}^3a_{ij} x_j(t)
\label{wcon}
\end{eqnarray}
where $x^i(t)=(x^{i}_{1}(t),x^{i}_{2}(t),x^{i}_{3}(t))^T\in R^3$,
$i=1,2,3,$ and $f(\cdot)$ is defined as in (\ref{Chua}).
\subsection{The coupling matrix is symmetric}
Suppose that the coupling matrix without the controller is
\[A= \left(
\begin{array}{ccc}
-5.1&5.0&0.1\\5.0&-11.0&6.0\\0.1&6.0&-6.1
\end{array}
\right )
\]
As for the coupling strength, we pick $c=10$.

Direct calculation shows that
\begin{equation}(x-y)^TP(f(x,t)-\Delta x-f(y,t)+\Delta y)
\leq -\eta (x-y)^T(x-y)
\end{equation}
where $P=I_{3}$, $\Delta=10 I_{3}$, $\eta=0.6218$.

Let the initial values be $x_1(0)=(40.1,20.2,30.3)^T$,
$x_2(0)=(20.4,30.5,10.6)^T$, $x_3(0)=(60.7,40.8,50.9)^{\top}$.
$s(t)=0$ is a solution of the uncoupled system.

Denote $x_{\xi}(t)=(x_1(t)+x_2(t)+x_3(t))/3$. We use the quantity
$\frac{\sum_{i=1}^{3}||x_{i}(t)-x_{\xi}(t)||}{\sum_{i=1}^{3}||x_{i}(0)-x_{\xi}(0)||}$
to measure synchronization capability. Figure 2 indicates that the
coupled system (\ref{wcon}) can reach synchronization without
controller. We also use the quantity
$\frac{\sum_{i=1}^{3}||x_{i}(t)-s(t)||}{\sum_{i=1}^{3}||x_{i}(0)-s(0)||}$
to measure if the specified solution $s(t)$ of the uncoupled
system is stable. Figure 3 indicates that $s(t)$ is unstable for
the coupled system (\ref{wcon}) without controller.

\begin{figure}
\centering
\includegraphics[width=2.5in]{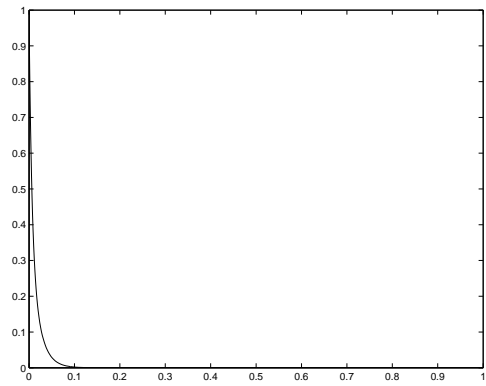}
\caption{The coupled system (\ref{wcon}) is synchronized without
controller.}
\begin{center} \label{fig_sim}
\end{center}
\end{figure}

\begin{figure}
\centering
\includegraphics[width=2.5in]{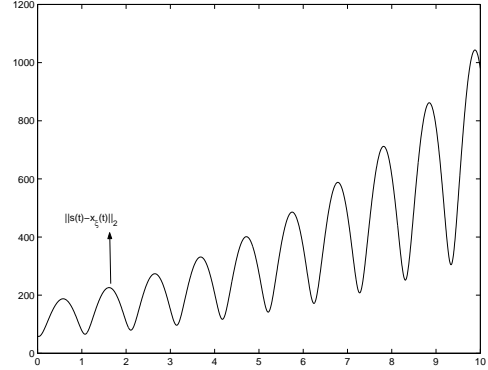}
\caption{The solution of coupled system (\ref{wcon}) does not
converge to s(t) without controller.}
\begin{center} \label{fig_sim}
\end{center}
\end{figure}

Now, we add a single controller to the first node of the coupled
system with $\varepsilon=4.9$
\begin{eqnarray}
\frac{dx_i(t)}{dt}=f(x_i(t),t)+c\sum\limits_{j=1}^3{}a_{ij}x_j(t)-\varepsilon(x_{1}(t)-s(t))
\nonumber\\ i=1,2,3 \label{withco}
\end{eqnarray}
Then, the coupling matrix becomes
\[\tilde{A}= \left(
\begin{array}{ccc}
-10.0&5.0&0.1\\5.0&-11.0&6.0\\0.1&6.0&-6.1
\end{array}
\right )
\]
The largest eigenvalue $\lambda_{1}=-10.11$ of the matrix
$\tilde{A}$. Therefore, $\Delta_k+c\lambda_1<0$ for
$k=1,\cdots,n$, and the conditions of Theorem 2 are satisfied. The
coupled system is synchronized to $s(t)$. Figure 4 shows
$\frac{\sum_{i=1}^{3}||x_{i}(t)-s(t)||}{\sum_{i=1}^{3}||x_{i}(0)-s(0)||}$.
The solution of coupled system (\ref{withco})  is pinned to s(t)
with a single controller

\begin{figure}
\centering
\includegraphics[width=2.5in]{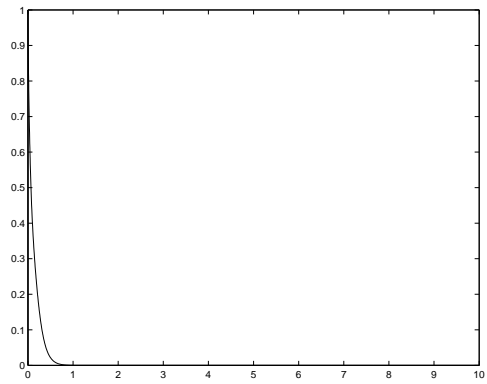}
\caption{The solution of coupled system (\ref{withco})  is pinned
to s(t) with a single controller.}
\begin{center} \label{fig_sim}
\end{center}
\end{figure}

\subsection{The coupling matrix is asymmetric}
In this simulation, we take the following asymmetric matrix
\[A= \left(
\begin{array}{ccc}
-2&1&1\\1&-2&1\\0&1&-1
\end{array}
\right )
\]
as the coupling matrix. As for the coupling strength, we pick
$\varepsilon=2$. So
\[\tilde{A}=\left(
\begin{array}{ccc}
-4&1&1\\1&-2&1\\0&1&-1
\end{array}
\right )\] Direct calculation indicates that the left eigenvector
corresponding to eigenvalue $0$ is
$\xi=(\frac{1}{6},\frac{2}{6},\frac{3}{6})^T$ and $\mu_1=0.0718$,
so we can choose $c=72$, such that theorem 4 is satisfied and the
simulation is shown by Figure 5.
\begin{figure}
\centering
\includegraphics[width=2.5in]{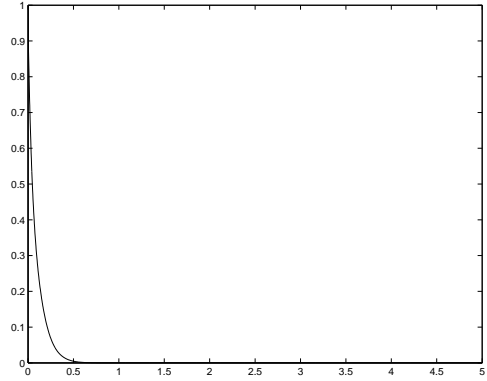}
\caption{The solution of coupled system (\ref{wcon}) converges to
s(t) through an asymmetric coupling}
\begin{center} \label{fig_sim}
\end{center}
\end{figure}

\section{Conclusions}
Synchronization is an important research field in sciences and
applications. How to pin a complex network to a specified solution
(or an equilibrium point) of the uncoupled system is of great
significance. However, in practice, the state variables of some
nodes are not observable or measured. Therefore, we have to
investigate the possibility of pinning a complex network by adding
controllers to those nodes, which can be measured or controlled.
In this paper, we prove rigorously that we can pin a complex
network by adding a single linear controller to one node with
symmetric or asymmetric coupling matrix. We also discuss how to
pin the complex network with nonlinear coupling. Simulations also
verify our theoretical results.

\end{document}